# The Last Meridian Circles – an Epilogue

Erik Høg
Niels Bohr Institute, Copenhagen University, Copenhagen, Denmark
ehoeg@hotmail.dk

**Abstract:** The aim is to document the last 35 years of meridian circles, a type of instrument with a fundamental role in astronomy for a very long time, and to do so while witnesses are still alive and can contribute. Meridian circles provided fundamental star positions for centuries. These positions were tied to a well-defined celestial coordinate system of right ascension and declination, and accurate proper motions ensured the transformation of the positions over long periods of time. This function of the meridian circles has been taken over by space astrometry and VLBI, Very Long Baseline Interferometry. The Hipparcos astrometric satellite was approved by ESA, the European Space Agency, in 1980 and launched a successful 3-year mission in 1989. Its successor, Gaia, completed a mission of 10.5 years in January 2025. An account is given of the last 18 meridian instruments, which were active for some part of the 35 years between 1980 and 2015. This account was made possible by input kindly supplied in correspondence with many colleagues.

## Introduction

The *meridian circle* (MC) was the fundamental instrument for astrometry for over two hundred years. Sometimes the term *transit circle* has been used for this type of instrument. It had first shown its great potential for measurement of star positions when Ole Rømer (1644-1710) set one up in 1704 near the village Vridsløsemagle, 20 km west of Copenhagen.[1] Advantages over the traditional quadrant were that it had an optical telescope on an axis resting its ends on two solid piers and had a full 360-degree circle with division lines which could be read and calibrated by means of several microscopes mounted in pairs at the end of a diameter.

Ole Rømer had first installed a *transit instrument* in his house in Copenhagen in 1691 (figure 1). The instrument had a scale (at F in the figure) which could be read with the microscope E. It was not yet a real meridian circle but both coordinates were determined, and both coordinates needed help from other instruments for the calibration. It was the first of its kind with a telescope mounted on an east-west axis so that the meridian transit time could be measured accurately and provide the right ascension. An accurate clock and an eyepiece micrometer were needed for the visual observations, and observations remained visual until photographic and photoelectric techniques were introduced after 1950.[2]

Ole Rømer's first real meridian circle is shown in figure 2. Another *classical meridian circle* of Ole Rømer's type from the 19$^{th}$ century is seen in figure 3. Both are suited for accurate measurement of right ascension and declination of stars.[3] A meridian circle of another design with full circle, a *mirror meridian circle*, is shown in figures 5 and 6.

Transit instruments in the meridian *without* an accurate circle were frequently used in the following centuries. With such instruments only the transit time and not the declination can be determined. This can provide the right ascension of stars, or just the clock time if the star has a known right ascension. Such an instrument is shown in figure 7. Finally, a transit instrument with full circle as in figure 7, a *vertical circle*, was used near the meridian to determine accurate declinations.

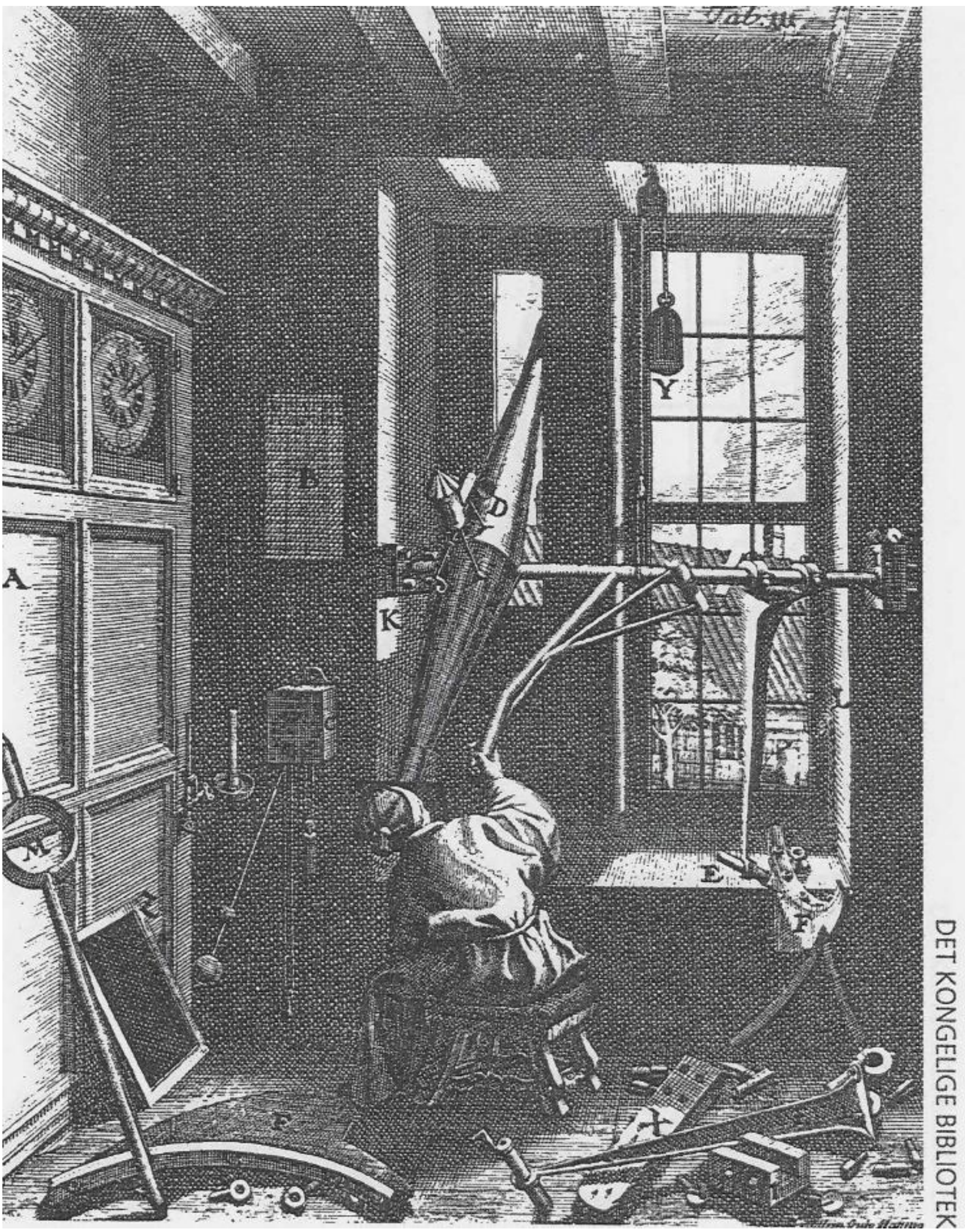

Figure 1. Engraving of Ole Rømer using his first transit instrument in 1691. Figure from Peder Horrebow, *Basis astronomiae sive Astronomiae pars mechanica : in qua describuntur observatoria, atque instrumenta astronomica Roemeriana Danica* (1735).

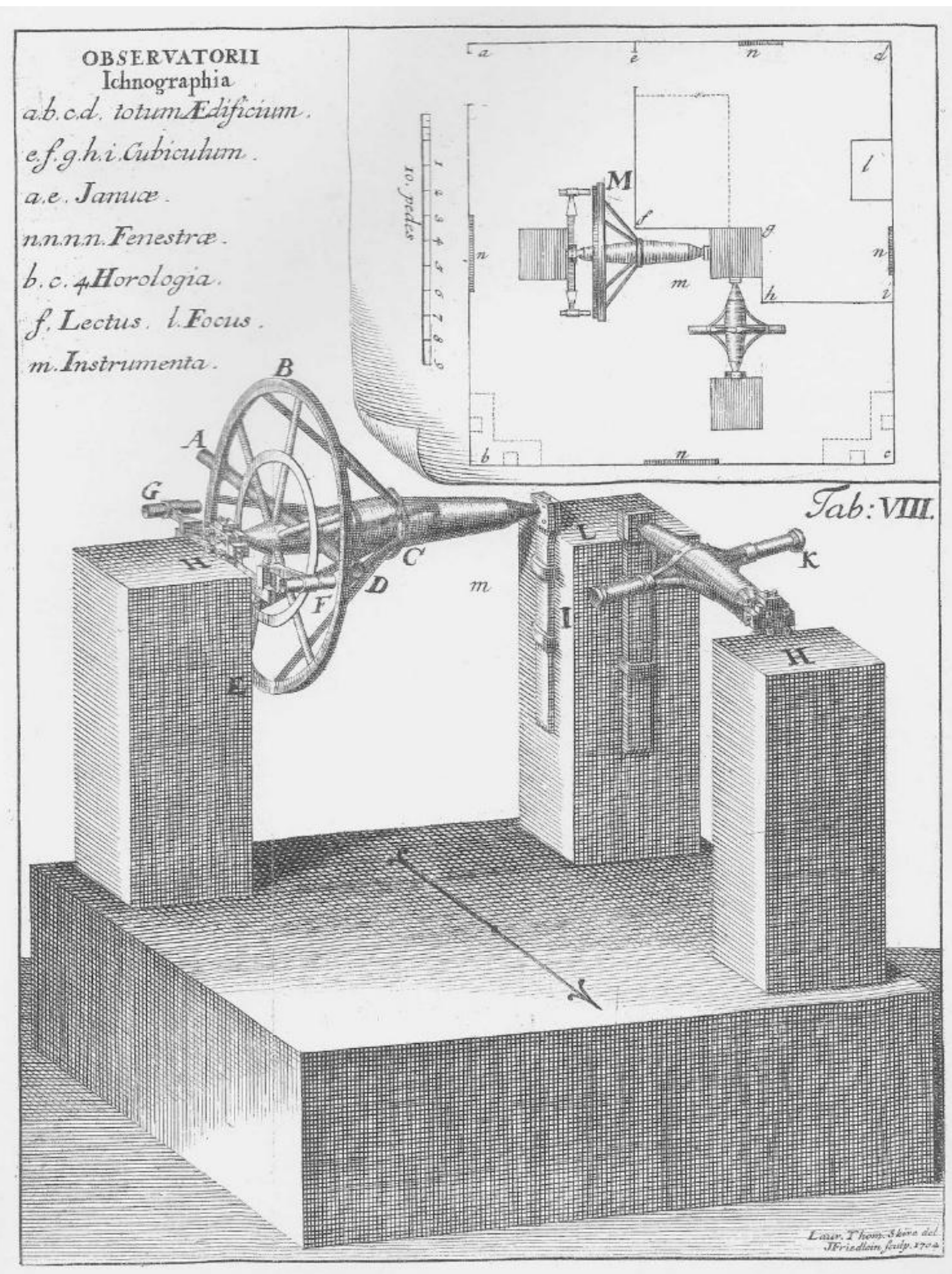

Figure 2. Engraving of Ole Rømer’s instruments of 1704. The apparatus at left is the meridian circle, which is marked M in the top view inserted at upper right. Axis of the meridian circle at H-L, telescope at A-D, circle at B-E, microscopes at G and F, and the inclination of the axis can be adjusted at I. Figure from Peder Horrebow, Basis Astronomiæ (1735).

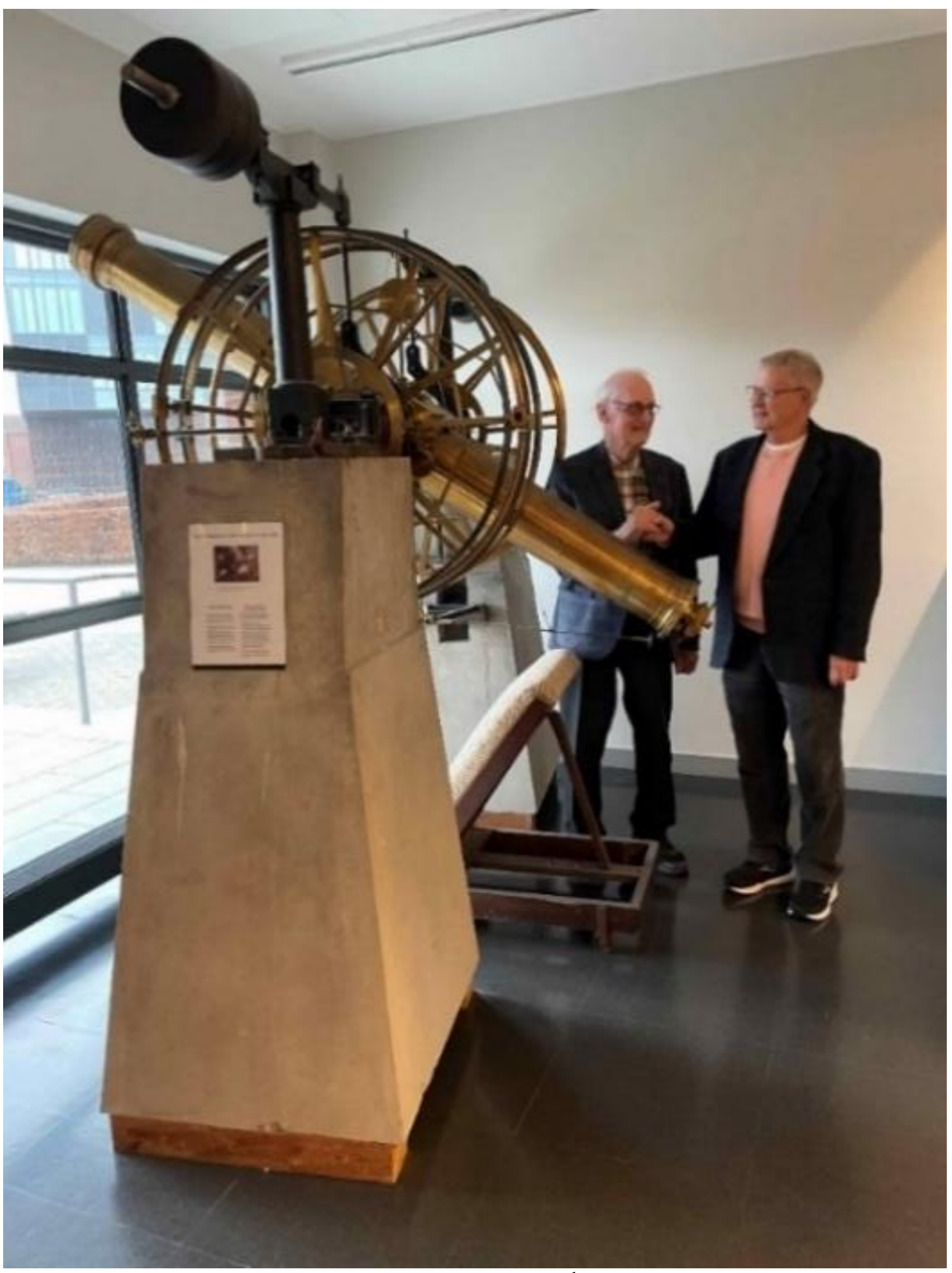

Figure 3. A classical meridian circle of 19th century design is now on permanent exhibition at Lund Observatory. Lennart Lindegren at left and the author are shaking hands in 2023. Astrometry and this instrument had fascinated Lennart in 1973, just as had happened for the author at the Brorfelde meridian circle 20 years before.[4] Figure from the private collection of Lennart Lindegren.

Observations with meridian circles served accurate determination of star positions during a long period, but such astrometric instruments became obsolete when ESA's astrometry satellite Hipparcos delivered very accurate positions, proper motions and parallaxes. The catalogue with 120,000 Hipparcos stars published in 1997 gave reference stars with 100 times smaller errors than the best ground-based positions. MCs could then be used to observe many fainter stars than Hipparcos, but MCs became obsolete also for this aim when the Gaia satellite reached out for 21st magnitude stars with a further 100 times better positions. The first results for 1.1 billion sources were published already in 2016; 10.5 years with Gaia observations of 1.8 billion stars were completed in January 2025; and the final results are expected in 2030.

The following were the locations of meridian instruments, not always meridian circles, that were active in at least a part of the 35 years from 1980 to 2015:

Brorfelde, Denmark
La Palma, Canary Islands. The same instrument as in Brorfelde
San Juan, Argentina
US Naval Observatory in Washington DC. 2 instruments
Flagstaff, Arizona
Mykolaiv (formerly Nicolaev), Ukraine
Kyiv University Observatory
The Main Astronomical Observatory of Kyiv
Moscow. 2 instruments
Kislovodsk. 2 meridian instruments
Pulkovo. 2 instruments
Tokyo
Bordeaux
Valinhos, Brazil
Cerro Calán, Chile

Of these 18 instruments, the six meridian circles in La Palma, San Juan, Flagstaff, Mykolaiv, Kyiv, and Bordeaux were still active after the year 2000.

**About methods and changing techniques**

A star is observed with a meridian circle while it crosses the meridian to obtain the right ascension and declination of the star. A micrometer at the focal plane of the telescope objective is used to measure the time of transit needed for the right ascension, and to measure the position in the field-of-view along the meridian needed for the declination. The inclination of the telescope during this measurement is determined from readings of a circle parallel to the meridian mounted on the east-west axis. The circle has very accurate division lines, and the reading is used with the measurement of the star along the meridian to finally give the celestial coordinate of declination. The small but significant error of each division line was determined in a separate calibration process. For calibration of the inclination of the east-west axis and of the zero-point of the circle a mercury pool often was placed at the nadir in which a reflected image could be observed several times in a night. Some meridian circles could be easily reversed 180 degrees around a vertical axis for further calibrations.

Stars were observed one by one, perhaps 200 in a good night, with manual setting of the telescope inclination before observation with the focal plane micrometer and reading of the circle. These stars could then by calculation be connected to a set of reference stars for which positions and motions were known from observation with meridian circles for more than 100 years. This was possible because a meridian circle is very stable during the night as it can only turn about an east-west axis resting in pivots on two solid columns. The net of reference stars after 1963 consisted of *1535 fundamental stars* on the whole sky listed in the FK4 catalogue[5], later replaced with the FK5[6], and the observer took care to observe a suitable number of them every night.

This situation changed completely with the publication of the Hipparcos Catalogue[7], much more accurate than FK4 and FK5 and with 100 times as many stars. And in 2000 the Tycho-2 Catalogue[8] appeared with data for 2.5 million stars also observed with Hipparcos.

In 1989 a new detector, a Charge Coupled Device (CCD) was being installed for astrometry at a transit telescope by Stone and Monet[9]. About 1994, CCD detectors were in use at some meridian circles so that many stars in a small area of the sky were observed simultaneously instead of only one star at a time. A strip of the sky was observed as the stars drifted across the CCD in so-called drift-scan mode (Time Delayed Integration, TDI) while the telescope stayed fixed. Drift-scan was also used in the Gaia satellite as proposed for an astrometric satellite called Roemer at a symposium in Shanghai in September 1992 by Erik Høg[10].

Some stars from Hipparcos or Tycho-2 were always present in the strip observed with the meridian circle to which the other stars in the strip could be connected. For such "relative astrometry", the great stability of a meridian circle during a night was no longer essential and an accurate reading of the circle was superfluous. Therefore, the name *meridian circle* was no longer fitting, and it was replaced by *meridian telescope* in some cases.

Furthermore, a CCD is much more sensitive than a photomultiplier so that even fainter stars can be observed. Some meridian circles or meridian telescopes were therefore dedicated to surveys of large parts of the sky for faint stars. That was the case for the instrument on La Palma until 2011 and similarly for some of the six meridian instruments in San Juan, Flagstaff, Nicolaev, Kiev, Tokyo, and Bordeaux. Thus, all meridian circles are decommissioned by 2015 because the Gaia mission does not leave significant scientific tasks for these instruments, and this is rather the same time as the involved persons retire.

**18 meridian instruments**

This is about the history of 18 instruments, their development, and their use and results as far as I know but far from complete. The period of activity after 1980 and the number of observed stars is mostly given for each instrument. Anecdotal information is included for lack of better but also to give the reader a flavor of the work and the relation among astronomers with whom I have been working for more than seventy years in astrometry, a field of research fundamentally changed in that period especially after approval of the Hipparcos mission by ESA in 1980.

The account is based on a study of the literature and on input received from many colleagues on my request, and I have edited what I received in close correspondence, resulting in different details and styles. I prefer to keep some details as they could be of historical interest as time passes. The intention is to tell as far as possible about the last years of a fundamental astronomical instrument but it is beyond my ability and not very important in the present context to provide all basic parameters on all the instruments—e.g., aperture of the meridian telescope, diameter of the circle and precision of the divisions, type and models of accessories such as photographic micrometers, latitude and longitude of meridian.

The 18 instruments are listed in a sequence with no special meaning; it came as the paper evolved with new information.

*1 Brorfelde*

The Copenhagen University Observatory, CUO, established a new Observatory near the village Brorfelde 50 km to the west of Copenhagen where the main instrument was a meridian circle built by Grupp Parsons in England and erected in 1953. A photographic micrometer was developed and in use 1964-1976.[11]

From 1973 a photoelectric slit micrometer with photon counting was developed, and full automation of the instrument was begun. Contact with the Royal Greenwich Observatory, RGO, in England from 1976 led to a collaboration where the automatic instrument should be set up on La Palma, one of the Canary Islands, where a large observatory was established in the following years on a mountain top, Roque de los Muchachos. Soon also Spain joined this collaboration. Observations with the MC on La Palma began in May 1984.[12]

*2 La Palma*

Leif Helmer, astronomer of the Copenhagen University Observatory, in 2025 informed me about the photoelectric observations:

"The observations on La Palma in the 14 years from May 1984 to May '98 were obtained with two photoelectric slit micrometers; the first one being replaced with a new one after four years. The latter micrometer had cooling of the photomultiplier, there were three pairs of inclined slits of different height (implemented by one pair of slits where a diaphragm could cover different parts) so that stars with uncertain positions could be measured, and there was a measuring rod engraved on the slit plate.

The results were published annually in *Carlsberg Meridian Catalogues*, Numbers 1 – 11, on CD-ROM in 1999. The entire catalogue contained 180,812 positions and magnitudes for 176,591 stars with declination > – 40º, 155,005 proper motions and 25,848 positions and magnitudes for 184 solar system objects. About 100,000 observations per year were obtained, so that the 181,000 positions are based on about 1.4 million observations with 8 observations/star.

A link from a Cambridge website by Evans[13] contains a reference to CMC14 of 2005. This is a catalog observed with the CCD in drift-scan mode covering the sky from −30 to +50 degrees in declination, just completed and published before Copenhagen University left the project and Leif Helmer retired in 2006. It contains astrometry and photometry catalogue of 95.9 million stars in the red (SDSS r′) magnitude range 9 to 17.

Finally comes the catalogue CMC15[14] which is the last of the series "Carlsberg Meridian Catalogue, La Palma" and comprises all the observations made between March 1999 and March 2011 with the Carlsberg Automatic Meridian Circle in El Roque de los Muchachos Observatory on the island of La Palma (Spain). It contains more than 122 million stars with right ascension, declination, and magnitude in the magnitude range $9 < r' < 17$ and declination range $-40° < \text{dec} < +50°$. [r′ is the magnitude in a particular red band.] For the catalogue, a mean value was formed from at least two observations per star, and the catalogue internal errors in astrometry are below 30 mas (milli-arcseconds) in both coordinates for stars brighter than r′=13, reaching 60 mas for r′=16. The internal magnitude error is below 0.020 mag for stars brighter than r′=13, and about 0.090 mag for r′=16.

This means almost 700 times (122 million/181,000) as many positions from the CCD-micrometers in 12 years as the 181,000 star-positions in 14 years from the slit micrometers. The last night of observing was 2013/09/01. The Carlsberg Meridian Telescope has been decommissioned, and the Instituto de Astrofisica de Canarias (IAC) has plans of housing the telescope in a museum in Santo Domingo de Garafía on La Palma, but this museum has nothing to do with IAC. The MC building on the mountain will be removed and the new Liverpool instrument will be located there."

*3 San Fernando/San Juan*

Real Instituto y Observatorio de la Armada en San Fernando (ROA) collaborated on the La Palma instrument and operated its own instrument in San Juan, Argentina. It was built by Grupp Parsons in 1948 as a twin to the one for Brorfelde and it was originally erected at San Fernando in 1952, according to Miguel Carrion.

Miguel Vallejo Carrion, astronomer of the San Fernando Observatory, informed me in 2025:

"Until 1986, it carried out numerous systematic visual observation campaigns of stars and objects of the solar system. In 1986 it was subject to an automation and robotization process, taking advantage of the collaboration started in 1980 with the RGO and CUO. In 1988 it was equipped with a photoelectric slit micrometer identical to the second one developed for the La Palma instrument.

When the automation process was finished in 1994 and after two years of testing in San

Fernando, the instrument was in ’96 transferred to the Carlos Ulrico Cesco (CUC) High Altitude Station, part of the Félix Aguilar Astronomical Observatory (OAFA) of the National University of San Juan (Argentina) thanks to a collaboration agreement between both institutions. The location is close to Leoncito, the American astrometric observatory.

At its new location, the instrument was renamed the Automatic Meridian Circle of San Fernando (CMASF) and was initially equipped with the slit micrometer (1997-1999). It carried out systematic observations of stars and small planets resulting in the publication in 2001 of the first Hispano-Argentinian Meridian Catalogue HAMC1 containing 6,192 positions and magnitudes of stars south of declination +40º, 5,400 proper motions and 923 positions and magnitudes of 92 objects in the solar system.

In 1999 a CCD camera replaced the slit micrometer and full automation was implemented, see Muiños.[15] A segment of the sky was observed between declinations −55º and +30º, but only the part south of equator was completed. This formed the Second Hispano-Argentinian Meridian Catalogue (HAMC2) published in 2008 containing more than 12,500,000 positions and magnitudes of stars between −30º and 0º in declination and magnitudes 9 < V < 16.5.

The instrument was finally closed down in 2014 due to staff limitations and the fact that results obtained from astrometric observations by the Gaia Space Mission made it scientifically unprofitable to continue with ground-based observations of lower precision.”

*4 USNO*

The US Naval Observatory (USNO) at Washington DC was a center for fundamental astrometry with meridian circles for many years. Other kinds of ground-based astrometry were also pursued at the USNO but are not the subject here. Plans for fundamental astrometry from space were studied in the 1990s in the USA.[16]

The meridian circles at USNO and the work with them for 150 years have been described by Schmidt and Dick,[17] who list the instruments in their Table 1. After 1980, two meridian circles were operated, the 6- and 7-inch Transit Circles, as quoted here:

“The Six-inch Transit Circle would become the true workhorse of fundamental astrometry at USNO in the 20th century, operating from 1899-1999 in the west transit building at the Massachusetts Avenue site in Washington, D.C. In that period, 974,000 visual observations were made with this instrument, 300,000 more than the Great Airy Transit Circle at Greenwich (1851-1954), perhaps more visual observations than any other transit circle in history.

The Seven-inch Transit Circle was the only meridian circle built at the Naval Observatory. It was nearly indistinguishable in design from its 50-years-older cousin, the Six-inch Transit Circle. But while the Six-inch remained a visual instrument, the Seven-inch pioneered electronic imaging automation. These instruments were managed for most of their existence by separate USNO Divisions, yet each participated in the global campaign of the 1950s-1960s known as the International Reference Stars, and in the USNO Pole-to-Pole Program of 1985-1996.

Theodore Rafferty made the final observations on the USNO Six-inch Transit Circle in 1999.”

*5 Flagstaff*

Located in Arizona, USA, the Flagstaff astrometric telescope (FASTT) has been described by Stone, *et al.*:[18]

“By the end of year 2003, the FASTT will have produced over 190,000 positions of solar system objects in a program to provide a very large and homogeneous database for each object that will extend over many years and include positions accurate to ±47 to ±300 mas, depending on the magnitude of each observed object ($3.5 < V < 17.5$). Moreover, extensive efforts have been undertaken to improve the systematic accuracy of FASTT equatorial positions by applying corrections in the reductions for differential color refraction, distortions in the focal plane, and correcting for a positional error that is dependent on magnitude. The systematic accuracy of FASTT observations is now about ±20 mas in both right ascension and declination. FASTT data have contributed very significantly to recent successful spacecraft missions and to a dramatic improvement in the predictions made for occultation events."

My comment: The FASTT at Flagstaff was a transit telescope observing stars close to the meridian. It was being upgraded with a CCD in 1989 (Stone & Monet 1990). A further upgrade started in 1997 and by 2003 a CCD with $2048^2$ pixels covered a field of 51′×51′. The positions were obtained by reference to the Tycho-2 Catalogue; thus the instrument was not working as a meridian circle where positions were derived by accurate observation of transit times and reading of the divided circle.

FASTT continued to take observations of the Saturnian satellites until at least 2007.[19]

Recently, Valeri Makarov, astronomer at the USNO, wrote that FASTT was decommissioned by 2010 after R.C. Stone had passed away in 2005.

*6 Mykolaiv*

In Ukraine, the Axial Meridian Circle (AMC) of the Mykolaiv (formerly Nicolaev) Astronomical Observatory is a horizontal rotating telescope (D = 180 mm, F = 2480 mm) located in the prime vertical, paired with a stationary long-focal-length autocollimator (D = 180 mm, F = 12,360 mm) and equipped with a CCD micrometer operating in drift-scan mode, figure 4.

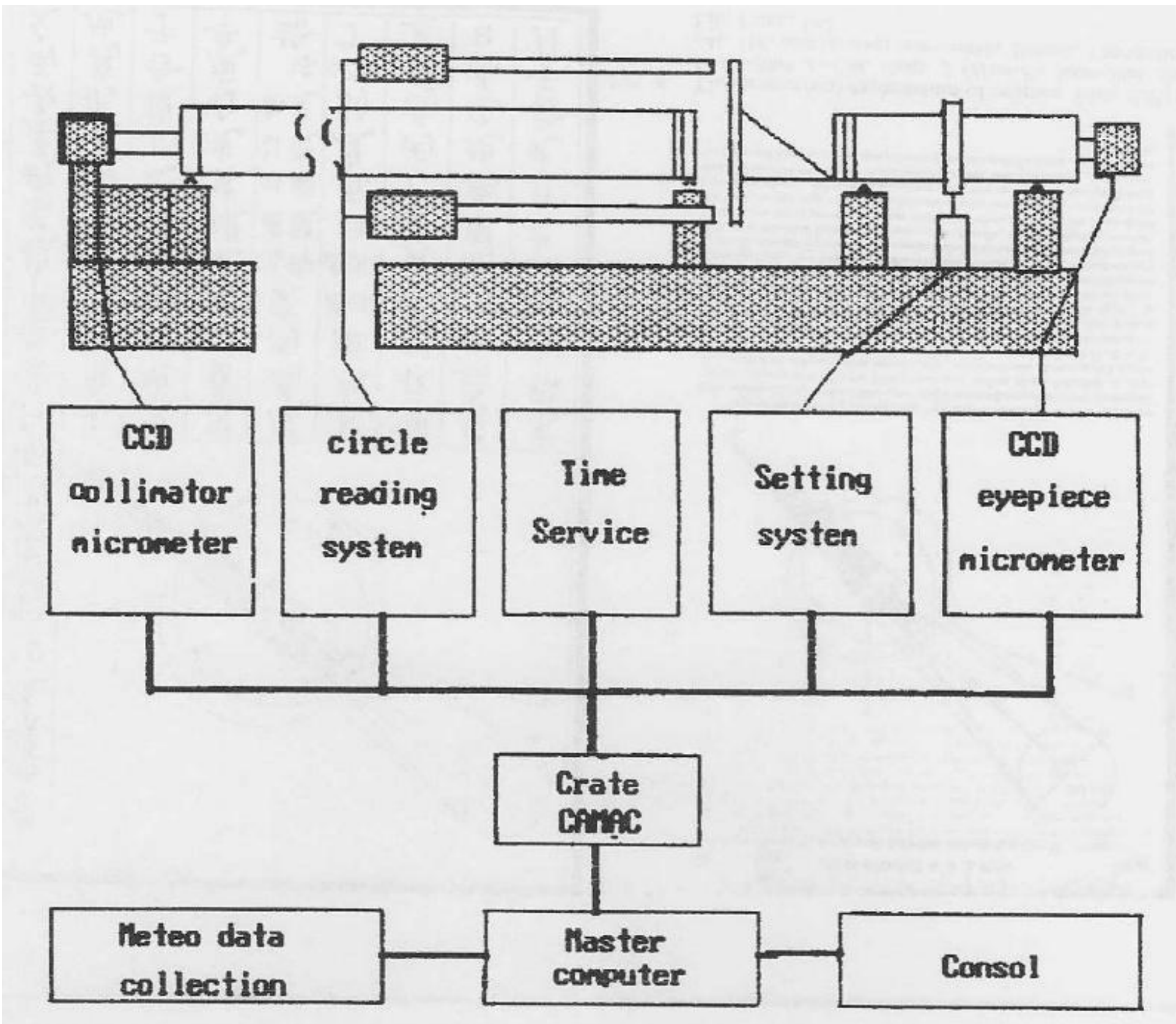

Figure 4. Operational structure of the Axial Meridian Circle (AMC) of the Mykolaiv Astronomical Observatory.[20]

Since 1995, the telescope has undergone a series of upgrades, including the replacement of the CCD receiver with a modern more sensitive one. The telescope was equipped with an Apogee Alta F8300 CCD camera featuring high-resolution sensors (8.3 megapixels with 5.4 µm pixels), providing a 25 × 25 arcmin field of view and a scale of 0.45 arcsec/pixel. Astrometry is entirely derived from reference stars within the field. Observation campaigns with the AMC have been focused on ground-based maintenance and extension of the Hipparcos reference system to fainter objects.[21]

Details of the AMC design have been given by Pinigin *et al.*[22] Observations from 2008–2011 are compiled in two catalogs with a total of 284,557 objects by Maigurova *et al.*[23] and Martynov *et al.*[24] Stars with large proper motions were observed by Maigurova and collaborators.[25]

Astrometric observations performed by the Gaia Follow-Up Network for Solar System Objects (Gaia-FUN-SSO) play a key role in ensuring that moving objects first detected by ESA’s Gaia mission remain recoverable after their discovery. An observation campaign on the potentially hazardous asteroid (99 942) Apophis was conducted during the asteroid’s latest period of visibility, from 12/21/2012 to 5/2/2013, to test the coordination and evaluate the overall performance of the Gaia-FUN-SSO.[26]

In 2015, observations were stopped in view of the much better results expected from the Gaia satellite.

*7 Kyiv-1*

From Ukraine, Liliya Kazanzeva, astronomer of the observatory, sent the following information in 2025:

“Our classical meridian circle was made by Repsold in 1870 and installed in Kyiv (formerly Kiev) Observatory 1872 in a special pavilion. It worked until 1996 in visual mode. When you arrived (the present author visited the meridian circle in 1978) you communicated with Chernega Mykola Yakimovich and he observed with this instrument for many years. Unfortunately, he has already passed away. Our observatory belongs to Kyiv University, but our employees also observed with the axial meridian circle at the other observatory in Kyiv [see the following section on Kyiv-2].”

*8 Kyiv-2*

From Ukraine, Yaroslav S. Yatskiv, director of the observatory, sent the following information in 2025:

“The Main Astronomical Observatory of Kyiv, belonging to the Academy of Sciences, had an axial meridian circle. It was constructed at the observatory and very similar to the one in Mykolaiv.

Observations were carried out with a CCD camera in the years 2001 to 2015.[27] They describe the results of the astrometric sky surveys with the telescope MAC which were performed in 2001-2015. They observed stars near the equator and stars in the fields with radio sources which are ICRF objects. Astrometry was entirely derived from reference stars within the field of view.

About a catalog with astrometric observations of 219,828 stars in the period 2001-2003 with the axial MC, see Lazorenko *et al.*[28] ”

*9 Moscow*

Valerian N. Sementsov, astronomer at Moscow State University, sent information in 2025 which is here abbreviated:

“Two meridian circles were installed and used in Sternberg Astronomical Institute, SAI, even after 1980:

1) Repsold’s meridian circle

(D=148 mm, F=2000 mm, manufactured in 1847).

It was originally installed in the Krasnopresnensky Observatory of Moscow State University (approximate coordinates 55.758046N, 37.567800E).

The meridian circle was moved in 1981 to the observatory on Maidanak Mountain in Uzbekistan (Google coordinates 38°40′19″N 66°53′55″E). In February 1993, all equipment at Maidanak was nationalized by the Uzbek authorities. Currently it belongs to the Ulugbek Astronomical Institute under the Academy of Sciences of Uzbekistan. Golovko mentions results[29].

2) Meridian circle of GOMZ (State Optical and Mechanical Plant) is shown in the undated figure at APM-4.[30] It shows the APM-4 - meridian circle with photographic registration (created under the direction of V.V. Podobed and A.P. Gulyaev) (D = 180 mm, F = 2500 mm, installed 1959-1961 in the pavilion with coordinates 55.701515N, 37.541488E).

Results are given by Tauber[31]. A catalogue of double stars' (DS) and high-luminosity stars' (HLS) declinations was obtained at the Moscow Observatory from observations with the meridian circle GOMZ. The observations were performed with the differential method by means of attachment to FK4 stars. The mean epoch of observations is 1982.70. The mean error of one observation is 0.30 arcsec."

The present author concludes that the two Moscow meridian circles were used up to about 1985, provided with visual or photographic micrometers, but no description seems to be available. This brief text is sufficient for the present purpose, and it has been agreed with by Oleg Malkov and Valerian N. Sementsov.

*10 Pulkovo/Kislovodsk*

In 1990, I visited the Pulkovo Observatory near Leningrad, later renamed St. Petersburg, invited to talk about the newly launched Hipparcos mission. My colleagues took me on a long trip, first to Moscow then to their observation station at Kislovodsk in the Caucasus Mountains. The visit to Kislovodsk had far reaching consequences for astrometry because I became engaged in discussions with my colleagues from Pulkovo: Mark Chubey, Valeri Makarov, and Vladimir Yershov, and soon also with Lennart Lindegren about a successor for the Hipparcos satellite. The discussions led directly to the Gaia mission.[32]

Vladimir N. Yershov, astronomer at the Pulkovo Observatory, in 2025 sent the following information about four astrometric instruments which were in use after 1980, two meridian circles at Pulkovo and two other types of meridian instruments at Kislovodsk, and he sent also a detailed illustrated paper that he had posted on the arXiv.[33]

*11 Pulkovo*

From correspondence with V. Yershov in January 2025:

"At Pulkovo, a Toepfer meridian circle was modernized with a new 200 mm objective and called MK-200.[34] It had a scanning photoelectric focal-plane micrometer as the one developed in Brorfelde by Erik Høg. A CCD-based photoelectric circle-reading system was built which allowed automatic reading of four microscopes, simultaneously with the run of the scanning focal-plane micrometer. All output and input was on punched tape.

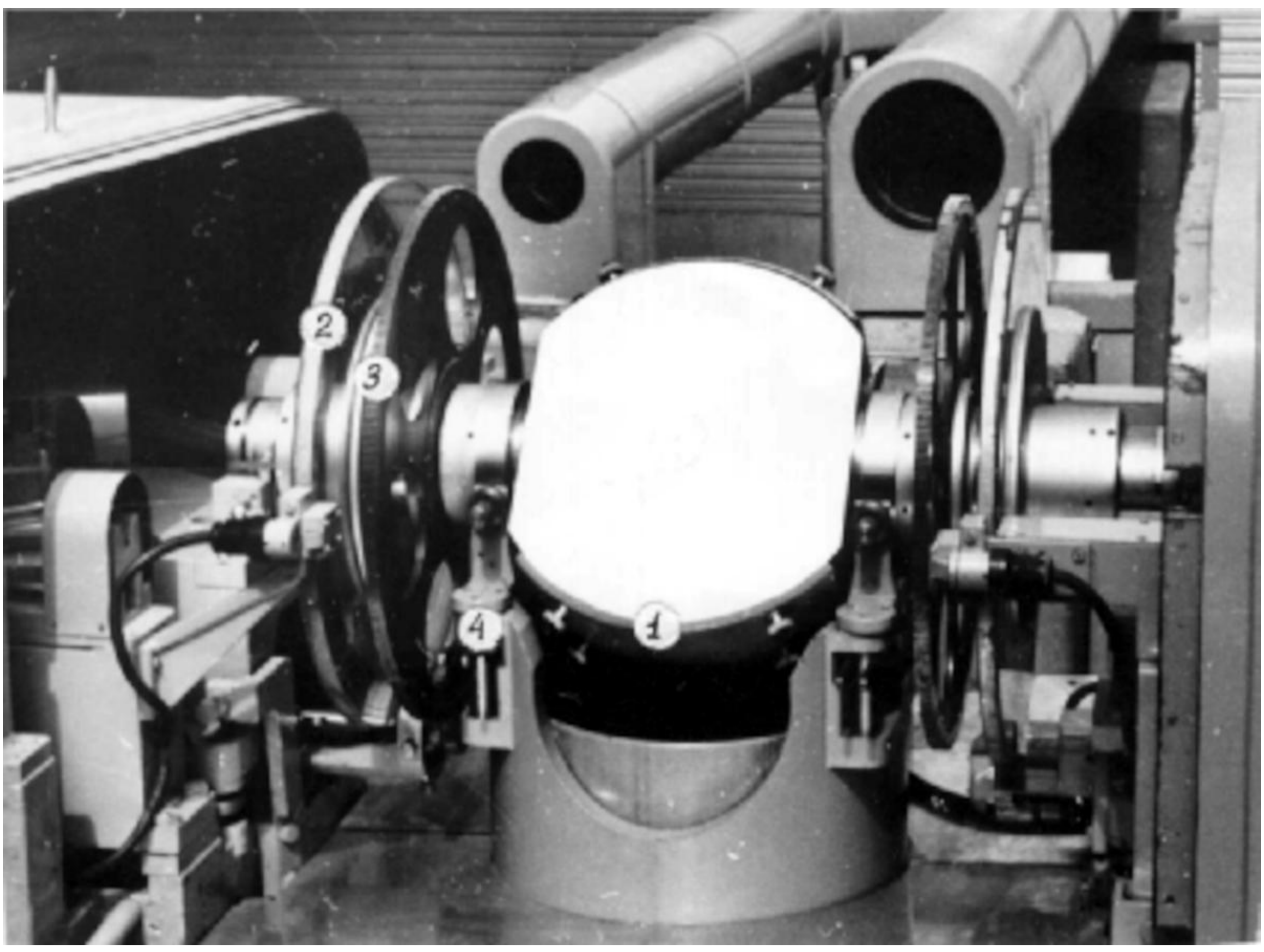

Figure 5. Sukharev Horizontal Meridian Circle of Pulkovo Observatory: 1- a flat mirror; 2- a graduated circle; 3 – a gear, 4 – an unloading system. Photo from about 1985 by the official Pulkovo photographer A.F. Sukhonos.

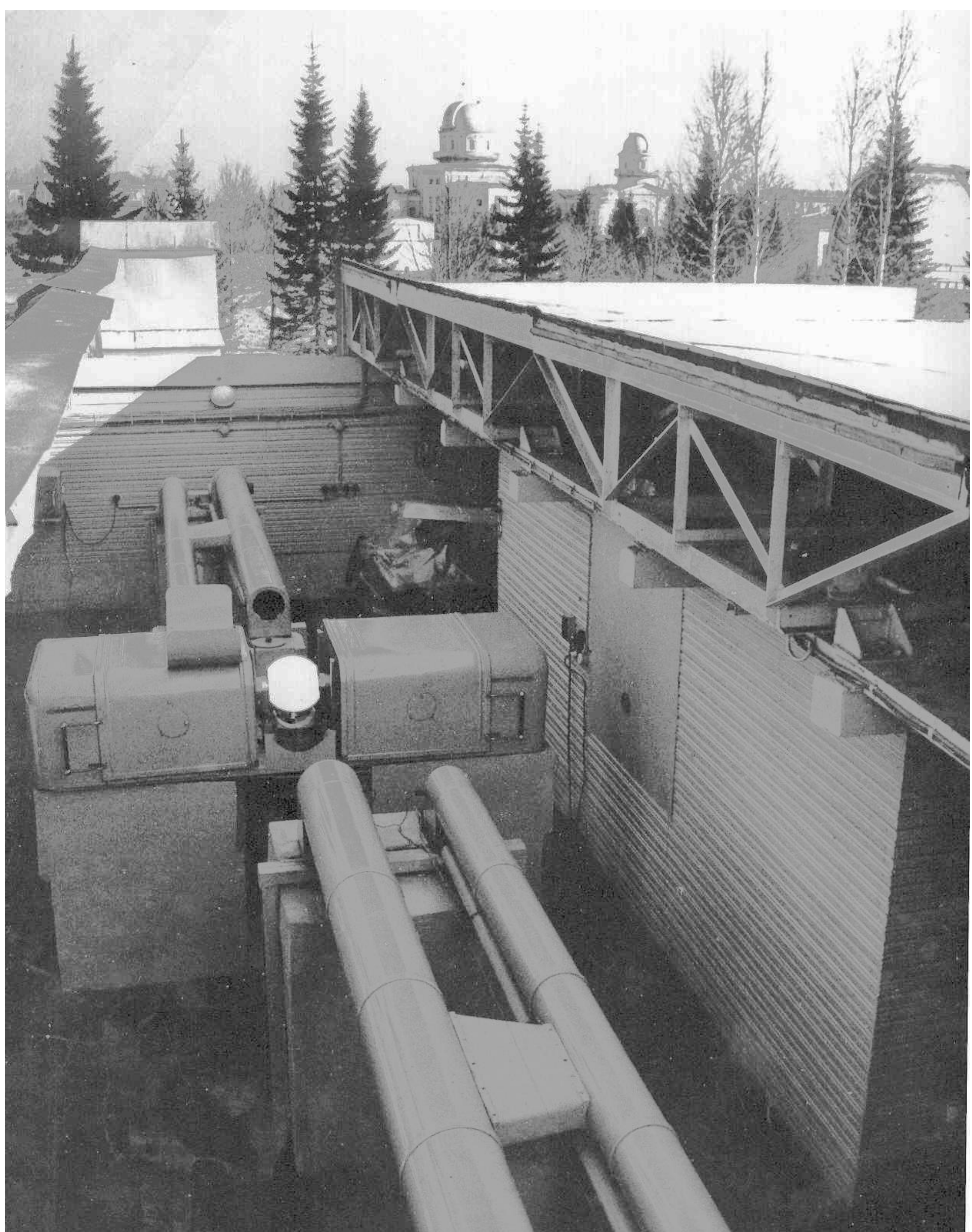

Figure 6. Flat-roof housing of the Sukharev Horizontal Meridian Circle. Photo from about 1985 by the official Pulkovo photographer A.F. Sukhonos.

Also at Pulkovo, a horizontal mirror meridian circle of special design by L.A. Sukharev was developed in the 1950s and after. See figures 5 and 6. In the 1980s, after development and study of an electronic control system, regular observations of bright and faint stars at both coordinates were carried out at the Sukharev Horizontal Meridian Circle in automatic mode.[35]

The researchers obtained a catalogue of positions of the faint part of the planned FK5, as well as a differential catalogue of positions of reference stars located around radio sources.

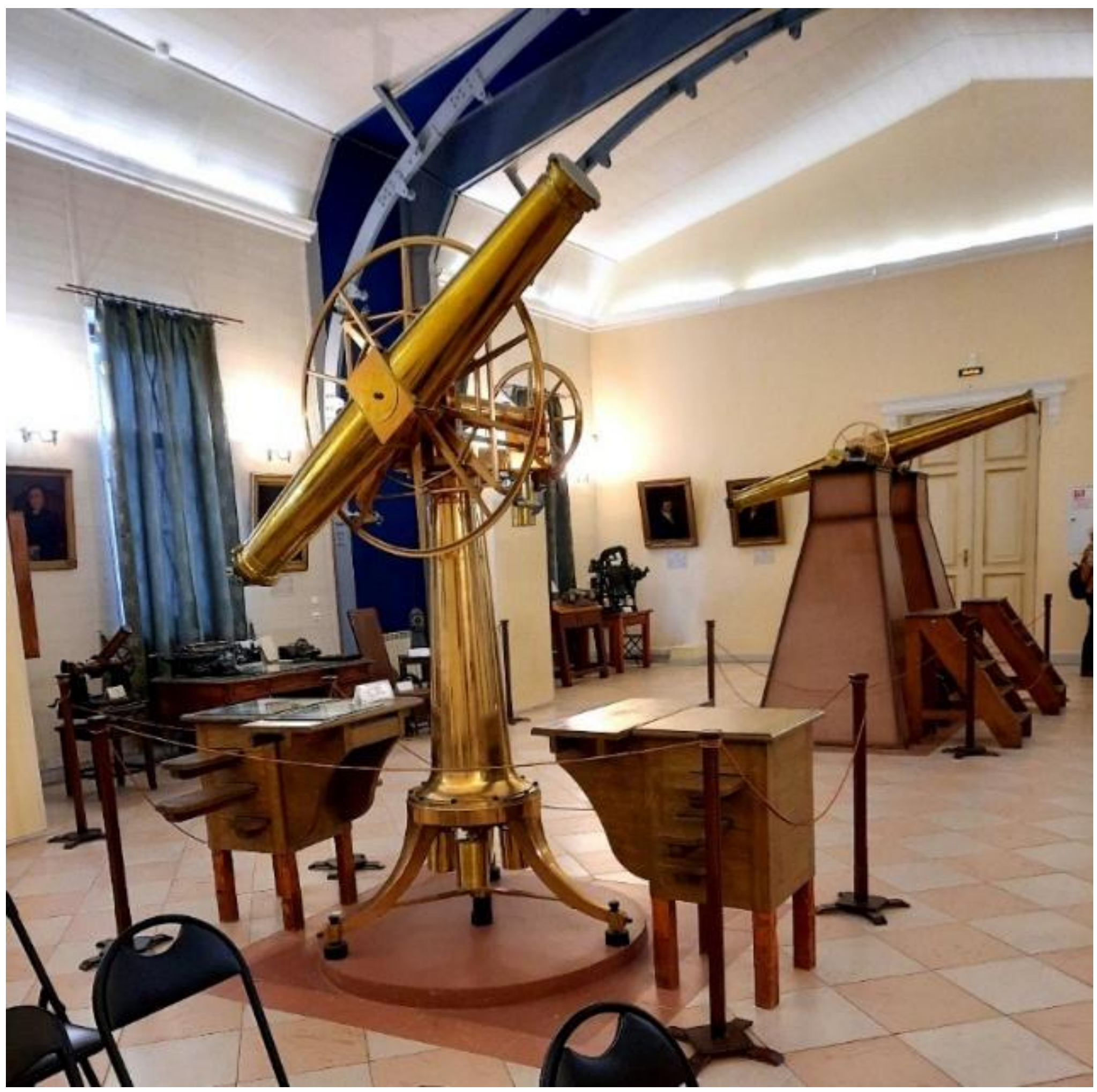

Figure 7. A classical Ertel-Struve vertical circle (in front) and a transit instrument (behind) at the Meridian Hall of Pulkovo Observatory (converted to an astrometry museum). Photo from 2024 by Vladimir N. Yershov.

At the Kislovodsk Mountain Station of Pulkovo Observatory there were no meridian circles. Instead, two classical instruments, a vertical circle and a transit instrument (see figure 7) designed in the 19th century by Struve and built by Ertel in Germany were installed in 1984 in two rhomb-shaped domes designed previously by Yu.S. Streletsky for a Large Transit Instrument installed at the Cerro-Calán National Astronomical Observatory (Santiago, Chile) in the 1960s-1970s. These two instruments were used for determining right ascensions and declinations of solar system bodies. The observations lasted from 1984 to 1989, and we have obtained 566 declinations of the Sun, 230 of Mercury, 413 of Venus, and 207 of Mars. Observations were visual, and the reductions were differential in the FK5 system."

Further observations from Kislovodsk were published by Devyatkin *et al.*[36] From their abstract: "A series of daytime observations [1983 to 1999] of the Sun and major planets are obtained at the mountain astronomical station of the Pulkovo Observatory using the Ertel-Struve meridian instruments. A series of declinations of Solar System bodies and major planets includes 4057 positions and that of right ascensions of Solar System bodies comprising 2057 positions. Based on the joint processing of observations of the Sun, Mercury, Venus, and Mars obtained with the Ertel-Struve vertical circle and large transit instrument."

*12 Tokyo*

The Tokyo Photoelectric Meridian Circle was a fully automated photoelectric meridian circle at the Mitaka campus of the National Astronomical Observatory of Japan in Tokyo, Japan. It was manufactured by Carl Zeiss Oberkochen, Germany and installed in Mitaka in 1982.[37]

Observations were published by Yoshizawa *et al.* about 3466 stars observed in 1989, and faint stars observed with CCD.[38] In a work of the following year[39], the results are presented from ten years of work, and the development of a new CCD micrometer and future observational plans are discussed. During the years 1985-93 a total of 199,275 observations of 35,300 stars were obtained with the scanning slit micrometer, furthermore 3684 observations of solar system objects, including the Sun. And a work from 1995 gives positions of Pluto.[40]

This message from Naoteru Gouda, astronomer at the observatory in Tokyo, was received in December 2024:

"Dear Erik Høg-san,
Hello.

I am very happy to hear that you are doing well.

Tokyo PMC is not currently in use, and observations ended in 2000. It is now open to the public and is playing an active role in outreach.

By the way, after Professor Masanori Miyamoto, who made an effort to build Tokyo PMC, retired at the end of fiscal year 1996, Yoshizawa-san was in charge of Tokyo PMC. I took up my post at the National Astronomical Observatory in fiscal year 1999 as Miyamoto-san's successor professor. However, I was not involved in Tokyo PMC, but instead started and focused on the study of a Japanese astrometry satellite mission. That is what led to the current JASMINE mission. JASMINE is moving towards realization. The Japanese government plans to launch JASMINE in fiscal year 2031. I appreciate your continued support.

I hope you have a happy new year.

Naoteru Gouda"

*13 Bordeaux and Valinhos*

The meridian circles are described by Viateau *et al.*[41]:

"A first CCD 512 × 512 camera working in scan mode (declination field 14′) was mounted in 1994 on the Bordeaux CCD meridian circle. After a testing period, this camera was installed on the Valinhos CCD meridian circle (near Sao Paulo, Brazil) as part of a collaboration between Bordeaux Observatory and the Instituto Astronomico e Geofisico of Sao Paulo. A second improved CCD 1024 × 1024 camera, with a declination field of 28′, was installed on the Bordeaux instrument in June 1996. The mean internal precision of a single observation is about 0.04″ in both coordinates for $9 \leq V \leq 14$."

Daniel Egret, former director of the Paris Observatory, has informed me: "I am sorry, but I have been unsuccessful, so far, in my attempts to get a full update about the status of the instrument.

A key point is the fact that the astronomy teams in Bordeaux have now left the historical site of the observatory. In 2017, the laboratory of astrophysics moved [to the University city site] and left the observatory site in Floirac. At that time, an association (Sirius) was created to counter the attempt by the university to sell the site. The sale project has since been abandoned but the site is almost abandoned: the instruments are starting to take on water...

Two of the last papers I have seen, beyond the observations of pre-main sequence stars together with the Mexican colleagues, are Ducourant *et al.*[42] with proper motions of 2.6 million stars and Arlot *et al.*[43] with observations of planets and satellites.

The meridian instrument of the Bordeaux observatory continued to produce observations until 2014."

*14 Valinhos*
This instrument is mentioned above, but we have no information about observations.

*15 Cerro Calán*
The Repsold meridian circle at Cerro Calán near Santiago de Chile was used for observations of major and minor planets in 1983 as reported by Carrasco and Loyola[44]. Carrasco *et al.* give a report about modernization of the instrument.[45]

**Conclusion: A Brief Status of Astrometry**
A few of the stunning results for *positions* from the Gaia mission should be mentioned but it would be too much here to give details also of the enormous output of proper motions, parallaxes, photometry, and radial velocities from the mission.[46] The accuracy of positions for 1.8 billion stars is better than 0.001 arcsec based on the first 33 months of Gaia observations and published in 2022. (See figure 8.) This is 100 times smaller errors than for the 1500 FK4 and FK5 stars, the most accurate catalogues for positions until the Hipparcos results appeared in 1997. The 17 million brightest stars in the sky have an accuracy of 0.014 milliarcsec, 10,000 times better than the positions in FK4/5. Orbits for 155,000 minor solar system bodies have been reconstructed. Gaia observations ended on 15 January 2025 after 10.5 years: 3827 days of science operations were concluded, 141,038 GB of science data gathered, and 267,336,261,480 transits observed.

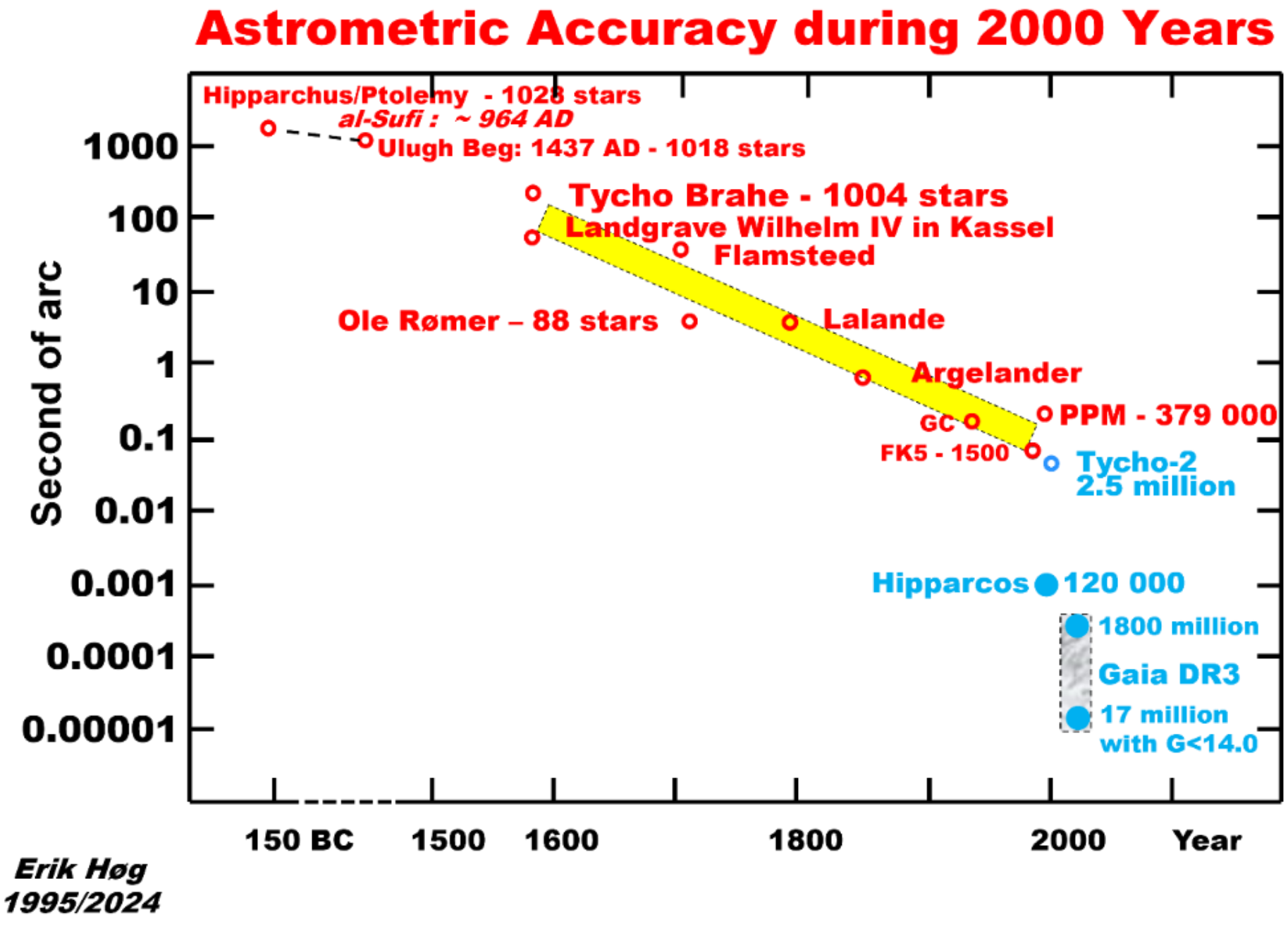

Figure 8. The accuracy of observed star positions for 2200 years. Observations until and including FK5 were obtained from the ground and the others from space. The accuracy was greatly improved shortly before 1600 AD by the Landgrave in Kassel and by Tycho Brahe in Denmark. The following 400 years brought an even larger but much more gradual improvement before space techniques with the European Hipparcos satellite started a new era of astrometry, with results denoted Tycho, Hipparcos, and Gaia in the figure. With Gaia, the 17 million brightest stars in the sky have accuracies of 0.014 milliarcsec from the first 33 months of observations, published as Gaia DR3. For G < 20.5 mag the accuracy is better than 1 milliarcsecond. Catalogs before 1900 are plotted at the mean epoch of observation, but after 1900 at the time of publication.[47]

On ground-based positions: The large drift-scans with meridian circles on La Palma and at San Juan resulted in positions and magnitudes for over 130 million stars. For solar system objects, 130,000 positions were obtained from La Palma, San Juan, and Flagstaff with accuracy about 0.1 arcsec. From Flagstaff is reported in 2003 that the astrometric data "have contributed very significantly to recent successful spacecraft missions and to a dramatic improvement in the predictions made for occultation events."

Meridian observations have been used in connection with Gaia. Results from the drift-scans were used to check for discontinuities in the Gaia photometry of the first Gaia catalogue in 2016, but not later, and the results were not used in connection with Gaia astrometry. Thuillot *et al.* wrote: "Astrometric observations performed by the Gaia Follow-Up Network for Solar System Objects (Gaia-FUN-SSO) play a key role in ensuring that moving objects first detected by ESA's Gaia mission remain recoverable after their discovery."[48]

I suppose that many people who had been working a long time to improve and use ground-based instruments may have had mixed feelings about Gaia. But I much hope and believe they will feel satisfied that they have dedicated their efforts the best way they could to a good scientific purpose: astrometry, and I hope they are pleased to see the immense progress obtained by space astrometry. Ground-based astrometry paved the way to this progress and took an active part to make it happen.

**Acknowledgements**
I am grateful for the information and comments from Daniel Egret, Dafydd Wyn Evans, Claus Fabricius, Naoteru Gouda, Leif Helmer, Liliya Kazanzeva, Lennart Lindegren, Valeri V. Makarov, Oleg Malkov, Valerian N. Sementsov, Catherine Turon, Miguel Vallejo Carrion, Yaroslav S. Yatskiv, and Vladimir N. Yershov. My special thanks go to an anonymous referee and the editor James Evans of *Journal for the History of Astronomy* for much help to improve and shape the paper.

**Notes on Contributor**
Erik Høg grew up on the Danish island Lolland, in the countryside under a dark sky full of stars. At the age of 14 he became interested in astronomy, built a reflecting telescope and observed variable stars. In 1954, he was given the task as a student at the Copenhagen University Observatory to make test observations with a new meridian circle. He became fascinated by the instrument and by the importance of astrometry for astronomy. He has been ever since deeply involved in the modern revolution of astrometry. This meant full digitization of meridian circle observations from 1960 at the Hamburg Observatory and later at Brorfelde. From 1975 leading roles in development of the two ESA astrometry satellites Hipparcos (including the Tycho experiment) and Gaia, launched in 1989 and 2013. In 2013 he proposed a successor for Gaia and this project, now led by David Hobbs, has a good chance to be launched by ESA about 2045.

**References**

---

[1] A.V. Nielsen, "Ole Rømer and his Meridian Circle", *Vistas in Astronomy*, 10 (1968), 105-112. https://ui.adsabs.harvard.edu/abs/1968VA.....10..105N/abstract ). C. Fabricius, N.T Jørgensen, and C.G. Tortzen, *Ole Rømer's Triduum*, Volume 1, *An introduction to the observations and a modern analysis*, and Volume 2, *Peder Horrebow's* Basis Astronomiæ (xxx: University Press of Southern

Denmark, 2023).

[2] On the changes to photographic and photoelectric techniques, see section 2.1 and section 4, respectively, of E. Høg, “A review of 70 years with astrometry”, *Astrophysics and Space Science*, 369 (2024), 23. https://rdcu.be/dzp2o with free access to the final manuscript at https://arxiv.org/abs/2402.10996

[3] Pictures of meridian circles are also found in Høg, “A review” (ref. 2), and in V.N. Yershov, “The Last Meridian Circles of Pulkovo Observatory”, https://arxiv.org/abs/2502.03583

[4] For more, see E. Høg, “Astrometriens udvikling og Brorfelde Observatorium,” *KVANT: Tidsskrift for Fysik og Astronomi*, bind 35, December 2024, 8-15. https://zenodo.org/records/14252551

[5] W. Fricke and A. Kopff, *Fundamental Catalogue (FK 4)*, Veröffentlichungen des Astronomischen Rechen-Instituts zu Heidelberg, Nr. 10 (Karlsruhe: Verlag Braun, 1963).

[6] W. Fricke, H. Schwan, T. Lederle, *et al.*, *Fifth Fundamental Catalogue (FK5)*, Part 1, *The Basic Fundamental Stars*. Veröffentlichungen des Astronomisches Rechen-Instituts Heidelberg, No. 32, (Karlsruhe: Braun, 1988). https://ui.adsabs.harvard.edu/abs/1988VeARI..32....1F/abstract Fricke, W., Schwan, H., Corbin, *et al.*, *Fundamental Catalogue (FK 5)*, Part 2, *The FK5 Extension - New Fundamental Stars*. Veröffentlichungen des Astronomisches Rechen-Instituts Heidelberg, No. 33 (Karlsruhe: Braun, 1991). https://ui.adsabs.harvard.edu/abs/1991VeARI..33....1F/abstract

[7] M.A.C. Perryman, L. Lindegren, J. Kovalevsky, *et al.*, “The HIPPARCOS Catalogue”, *Astronomy and Astrophysics*, 323 (1997), p. L49-L52.

[8] E. Høg, “Astrometry and Photometry of 400 million Stars Brighter than 18 Mag”, *Developments in astrometry and their impact on astrophysics and geodynamics: proceedings of the 156th Symposium of the International Astronomical Union* (Kluwer Academic: Dordrecht, 1993), p. 37. https://articles.adsabs.harvard.edu/pdf/1993IAUS..156...37H E. Høg, C. Fabricius, V. Makarov, *et al.*, “The Tycho-2 catalogue of the 2.5 million brightest stars”, *Astronomy and Astrophysics*, 355 (2000), p. L27-L30.

[9] R.C. Stone and D.G. Monet, 1990, “The USNO Flagstaff Station CCD Transit Telescope and Star Positions Measured from Extragalactic Sources. Inertial Coordinate System on the Sky”, *Proceedings of IAU Symposium No. 141 held 17-21 October 1989 in Leningrad, USSR*, ed. by J.H. Lieske and V.K. Abalakin (Dordrecht: Kluwer Academic, 1990), p. 369. https://articles.adsabs.harvard.edu/full/1990IAUS..141..369S

[10] E. Høg, “Astrometry and Photometry”, (ref. 8).

[11] On the development of astrometry at the Brorfelde Observatory see Høg, “Astrometriens udvikling og Brorfelde Observatorium” (ref. 4).

[12] See figure 10 in Høg, “A review”, (ref. 2).

[13] D.W. Evans, The Carlsberg Meridian Telescope (2014), https://research.ast.cam.ac.uk/cmt/camc.html

[14] J.L. Muiños and D.W. Evans, "The CMC15, the last issue of the series "Carlsberg Meridian Catalogue, La Palma", *Astronomische Nachrichten*, 335 (2014), no. 4, p. 367. https://ui.adsabs.harvard.edu/abs/2014AN....335..367M/abstract

[15] See J.L. Muiños, *et al.*, The San Fernando Automatic Meridian Circe in San Juan" (1999). https://people.ast.cam.ac.uk/~dwe/AstSurv/pepe.html

[16] See, e.g., K.J. Johnston, P.K. Seidelmann, *et al.*, "Newcomb Astrometric Satellite", *Astronomical and astrophysical objectives of sub-milliarcsecond optical: astrometry: proceedings of the 166th Symposium of the International Astronomical Union* (Dordrecht: Kluwer Academic, 1995), 331 https://ui.adsabs.harvard.edu/abs/1995IAUS..166..331J/abstract and the overview in E. Høg, "Interferometry from Space: A Great Dream", *Asian Journal of Physics*, 23 (2014), nos. 1 and 2, Special Issue on History of Physics and Astronomy, ed. Virginia Trimble http://arxiv.org/abs/1408.4668

[17] R.E. Schmidt and S.J. Dick, "Meridian Circle Astrometry at the U. S. Naval Observatory", *Cahiers François Viète* [En ligne], III-14 | 2023, mis en ligne le 01 juin 2023. http://journals.openedition.org/cahierscfv/4086 ; DOI: https://doi.org/10.4000/cahierscfv.4086

[18] R.C. Stone, D.G. Monet, A.K.B. Monet, *et al.*, "Upgrades to the Flagstaff Astrometric Scanning Transit Telescope: A Fully Automated Telescope for Astrometry", *The Astronomical Journal*, 126 (2003), 2060-2080. https://iopscience.iop.org/article/10.1086/377622

[19] See J. Desmars, A. Vienne, and J.-L. Arlot, "A new catalogue of observations of the eight major satellites of Saturn (1874–2007)", *Astronomy & Astrophysics*, 493 (2009), 1183–1195. https://www.aanda.org/articles/aa/pdf/2009/03/aa10203-08.pdf

[20] Diagram from G. Pinigin, *et al.*, "The axial meridian circle of the Nikolaev astronomical observatory", *Astronomical & Astrophysical Transactions*, 8:2 (1995), 161-163. https://images.astronet.ru/pubd/2008/09/28/0001230037/161-163.pdf

[21] A. Kovalchuk, *et al.*, "Mykolayiv AMC positions of faint stars in selected fields around extragalactic radio sources for linking the optical and radio reference frames", *The New International Celestial Reference Frame*, 23rd meeting of the IAU, Joint Discussion 7, 22 August 1997, Kyoto, Japan, meeting abstract id.16. https://ui.adsabs.harvard.edu/abs/1997IAUJD...7E..16K/abstract A. Kovalchuk, *et al.*, "Extension of the Hipparcos catalogue to the faint stars in the selected fields using the Nikolaev telescope AMC", *Motion of Celestial Bodies, Astrometry and Astronomical Reference Frames* (Proceedings of Journées 1999 and IX Lohrmann-Kolloquium, Dresden, September 13-15, 1999), ed. by N. Capitaine and M.H. Soffel (Paris: Observatoire de Paris, 2000), p. 146, no abstract. https://ui.adsabs.harvard.edu/abs/2000jsrs.meet..146K/abstract

[22] G.I. Pinigin, A.N. Kovalchuk, Y.I. Protsyuk, and A.N. Shulga, 2001, "Automatic Axial Meridian Circle of Nicolaev Astronomical Observatory", *Astronomische Gesellschaft Abstract Series*, 18 (2001), abstract #P231. https://ui.adsabs.harvard.edu/abs/2001AGM....18.P231P/abstract

[23] N.V. Maigurova, M.V. Martynov, G.I. Pinigin, "Star catalogue in the fields of axial meridian circle of the Nikolaev observatory", *Kinematics and Physics of Celestial Bodies*, 29 (2013), no. 1, 44-51. https://ui.adsabs.harvard.edu/abs/2013KPCB...29...44M/abstract

[24] M.V. Martynov, N.V. Maigurova, G.I. Pinigin, "Astrometric binaries with invisible components in fields of the axial meridian circle of the Nikolaev Observatory", *Kinematics and Physics of Celestial Bodies*, 29 (2013), no. 5, 254-256. https://ui.adsabs.harvard.edu/abs/2013KPCB...29..254M/abstract

[25] N.V. Maigurova, M.V. Martynov, and V.F. Kryuchkoskiy, "Results of astrometric observations of stars with large proper motions using telescopes of the Research Institute of Nikolaev Astronomical Observatory", *Kinematics and Physics of Celestial Bodies*, 32 (2016), no. 6, 307-312. https://ui.adsabs.harvard.edu/abs/2016KPCB...32..307M/abstract

[26] W.Thuillot, D. Bancelin, A. Ivantsov, *et al.*, "The astrometric Gaia-FUN-SSO observation campaign of 99942 Apophis", *Astronomy & Astrophysics*, 583 (2015), id.A59, 12 pp. https://ui.adsabs.harvard.edu/abs/2015A%26A...583A..59T/abstract

[27] See the following two papers. V.L. Karbovsky, P.F. Lazorenko, V.N. Andruk, *et al.*, "Kyiv meridian axial circle with a new CCD camera", *Kinematics and Physics of Celestial Bodies*, 27 (2011), no. 4, 204-210. https://ui.adsabs.harvard.edu/abs/2011KPCB...27..204K/abstract V.L. Karbovsky, P.F. Lazorenko, *et al.*, "The programs of observations on MAC in 2001-2015 and their results", *Bulletin of Taras Shevchenko National University of Kyiv*, no. 55 (2017), 17-19. https://ui.adsabs.harvard.edu/abs/2017BTSNU..55...17K/abstract

[28] P.F. Lazorenko, Yu. Babenko, V.L. Karbovsky, *et al.*, "The Kyiv Meridian Axial Circle Catalogue of stars in fields with extragalactic radio sources", *Astronomy & Astrophysics* (2018) manuscript no. 2573, 13 pp. https://arxiv.org/pdf/astro-ph/0512533

[29] B.A. Golovko, A.P. Gulyaev, V. G. Tauber, V.G. Shamaev, "Results of the First Observations with a Meridian Circle in Mountain Conditions – Part two – Processing of right ascensions", *Astronomicheskii Tsirkulyar*, No. 1304/Jan 05, 1984, p.1. https://ui.adsabs.harvard.edu/abs/1984ATsir1304....1G No abstract is given.

[30] An undated photograph of the APM-4 meridian circle with photographic registration may be seen at https://www.astronet.ru/db/msg/1176727/mer_krug.jpg.html

[31] V.G. Tauber, "The Catalogue of the Declinations of the DS and Hls Programme 513 Stars", *Trudy Gosudarstvennogo Astronomicheskogo Instituta P.K. Sternberga*, 58 (1986), p. 146. https://ui.adsabs.harvard.edu/abs/1986TrSht..58..146T

[32] See section 14 of Høg, "A review" (ref. 2).

[33] Yershov, "The Last Meridian Circles of Pulkovo Observatory" (ref. 3).

[34] See V.N. Ershov, V.E. Pliss, Y.S. Streletsky, "The Photoelectric Meridian Circle of the Pulkovo", *Astrometric Techniques*: *Proceedings of IAU Symposium No. 109 held 9-12 January 1984 in Gainesville, FL*, ed. by H.K. Eichhorn and R.J. Leacock (Dordrecht: D. Reidel Publishing Co., 1986), 407-411. https://ui.adsabs.harvard.edu/abs/1986IAUS..109..407E/abstract

[35] T.R. Kirian and G.I. Pinigin, "First results of observations of star declinations on the Pulkovo horizontal meridian circle", *Astronomicheskii Zhurnal*, 60 (1983), 775-780. *Soviet Astronomy*, 27 (1983), 448-451. https://ui.adsabs.harvard.edu/abs/1983AZh....60..775K/abstract

[36] A.V. Devyatkin, K.G. Gnevysheva, G.D. Baturina, "Results of astrometrical observations of the Sun and major planets at the mountain astronomical station of the Pulkovo Observatory", *Solar System Research*, **43** (2009), 533–542. https://doi.org/10.1134/S0038094609060082

[37] See M. Yoshizawa, M. Miyamoto, M. Sôma, T. Kuwabara, and S. Suzuki, "Ten years of the Tokyo Photoelectric Meridian Circle (Tokyo PMC): an activity report for the years from 1982 to 1993", *Publications of the National Astronomical Observatory of Japan*, 3 (1994), 289-301 https://ui.adsabs.harvard.edu/abs/1994PNAOJ...3..289Y/abstract

[38] M. Yoshizawa and S. Suzuki, "The Tokyo PMC catalog 89: catalog of positions of 3466 stars observed in 1989 with Tokyo Photoelectric Meridian Circle", *Publications of the National Astronomical Observatory of Japan*, 3 (1993), 45-91. https://ui.adsabs.harvard.edu/abs/1993PNAOJ...3...45Y/abstract M. Yoshizawa, S. Suzuki, T. Kuwabara, and H. Ishizaki, H. 1993, "Observations of Faint Stars Deep to 16TH Magnitude with CCD Meridian Circle," *Developments in astrometry and their impact on astrophysics and geodynamics: proceedings of the 156th Symposium of the International Astronomical Union held in Shanghai; China; September 15-19; 1992.* (Kluwer Academic Publishers, Dordrecht, 1993), 71. https://ui.adsabs.harvard.edu/abs/1993IAUS..156...71Y/abstract

[39] Yoshizawa, *et al.*, "Ten years of the Tokyo Photoelectric Meridian Circle" (ref. 37).

[40] M. Yoshizawa, M. Soma, S. Suzuki, "Positions of Pluto in 1994 Observed with the Tokyo CCD Meridian Circle", *Astronomical Journal*, 110 (1995), 3050-3053. https://ui.adsabs.harvard.edu/abs/1995AJ....110.3050Y/abstract

[41] B. Viateau, Y. Requieme, J.F. Le Campion, *et al.* 1999, "The Bordeaux and Valinhos CCD meridian circles", *Astronomy and Astrophysics Supplement Series*, 134(1) (1999). DOI: 10.1051/aas:1999131. https://www.researchgate.net/publication/44844082_The_Bordeaux_and_Valinhos_CCD_meridian_circles

[42] C. Ducourant, J.F. Le Campion, M. Rapaport, M., *et al.*, 2006, "The PM2000 Bordeaux proper motion catalogue", *Astronomy & Astrophysics*, 448 (2006), 1235-1245. This is a proper motion catalogue of 2 670 974 stars, covering the declination zone +11 to +18°. https://www.aanda.org/articles/aa/full/2006/12/aa3220-05/aa3220-05.html

[43] J.E. Arlot, G. Dourneau, and J.F. Le Campion, "An analysis of Bordeaux meridian transit circle observations of planets and satellites (1997–2007)", *Astronomy & Astrophysics*, 484 (2008), 869-877. https://www.aanda.org/articles/aa/abs/2008/24/aa9166-07/aa9166-07.html

[44] G. Carrasco, and P. Loyola, "Observaciones de planetas y pequeños planetas realizadas en 1983, en el círculo meridiano Repsold de Cerro Calán" *Boletín de la Asociación Argentina de Astronomía*, 30 (1984), 225-230. https://ui.adsabs.harvard.edu/abs/1984BAAA...30..225C/abstract

[45] G. Carrasco, P. Loyola, and N. Hadad, 1987, "Automatización de las observaciones en ascensión recta, en el Círculo Meridiano Repsold de Cerro Calán", *Boletín de la Asociación Argentina de Astronomía*, 32 (1987), 289-294. https://ui.adsabs.harvard.edu/abs/1987BAAA...32..289C/abstract

[46] ESA Gaia space mission https://www.esa.int/Science_Exploration/Space_Science/Gaia

[47] Figure from E. Høg, "Astrometric Accuracy of Positions" (2024), 8pp. https://arxiv.org/abs/2405.02017

[48] Thuillot, *et al.*, "The astrometric Gaia-FUN-SSO observation campaign" (ref. 26).